\RequirePackage{ifpdf}
\ifpdf 
\documentclass[pdftex]{sigma}
\else
\documentclass{sigma}
\fi

\begin{document}

\allowdisplaybreaks

\renewcommand{\PaperNumber}{105}

\FirstPageHeading

\ShortArticleName{Noncommutative Root Space
 Witt, Ricci Flow, and Poisson Bracket}

\ArticleName{Noncommutative Root Space
 Witt, Ricci Flow,\\ and Poisson Bracket Continual Lie
Algebras}

\Author{Alexander ZUEVSKY}

\AuthorNameForHeading{A. Zuevsky}

\Address{School of Mathematics, Statistics and Applied Mathematics,
National University of Ireland,\\ Galway, Ireland}

\Address{Max-Planck Institut f\"ur Mathematik, Vivatsgasse 7, 53111, Bonn, Germany}

\Email{\href{mailto:zuevsky@mpim-bonn.mpg.de}{zuevsky@mpim-bonn.mpg.de}}

\ArticleDates{Received August 21, 2009, in f\/inal form November 16, 2009;  Published online November 19, 2009}

\Abstract{We introduce new examples of mappings
 def\/ining noncommutative root space generalizations
 for the Witt, Ricci f\/low, and Poisson bracket continual Lie algebras.}

\Keywords{continual Lie algebras; noncommutative
integrable models}

\Classification{35Q58; 37K05; 37K30}

\section{Introduction}

The notion of a continual Lie algebra (a Lie algebra with continual set
of roots) f\/irst appeared in works of Saveliev and Vershik
\cite{s10, sv2, sv3, s}.
 The main feature which distinguishes continual Lie algebras from
 ordinary ones is that their generators are parametrized
by elements of a~space which we call in what follows
the {\it root space} ${\cal E}$.
The commutator of two continual Lie algebra elements also depends on bilinear mappings $K$
def\/ined on the root space.
Continual Lie algebras~\cite{sv2} represent generalizations of
many classes of inf\/inite-dimensional Lie algebras.
In the simplest class of continual Lie algebra examples,
Kac--Moody Lie algebras \cite{kac}, mappings $K$ reduce to Cartan matrices
corresponding to discrete root spaces.

The general def\/inition of a continual Lie algebra \cite{sv2, sv3}
 admits a noncommutative space of roots.
 The identities which follow from Jacobi identity
applied to elements of a continual Lie algebra
are also valid in noncommutative case.
Nevertheless, it is quite complicated to f\/ind examples (some
of them were derived in \cite{zu0}).
 Most cases of mappings def\/ining continual Lie algebras
 with commutative root spaces do not satisfy to these
identities when one passes to a~noncommutative space.
It is even more complicated when original mappings contain derivative terms.
 Our aim was to f\/ind classes of appropriate dif\/ferential mappings
 that are subject to the identities (\ref{redjac1})--(\ref{redjac4})
when ${\cal E}$ is noncommutative.

In this paper we introduce generalizations for the Witt \cite{ds}, Ricci f\/low
\cite{bakas1, bakas2, bakas3, zuz2}, and Poisson bracket~\cite{sv2},
as well as other examples of  continual Lie algebras with noncommutative root spaces.
 The noncommutative root space ${\cal E}$ we use is the space of
tensor product powers of a~space~$E$ which is endowed with a noncommutative product.
We also introduce some special products def\/ined in~${\cal E}$.
The associativity of these products leads to solutions
of the identities \mbox{(\ref{redjac1})--(\ref{redjac4})}.
We would like to stress that we derive mappings for new continual Lie algebras
independently of the nature of a noncommutative product in~$E$.
In further applications one can then consider examples of continual Lie algebras with
some specif\/ic multiplications  in $E$ (as in, e.g., see~\cite{lich}), and with some
f\/ixed commutation relations of $E$ elements and their derivatives
with respect to parameters.
New examples of continual Lie algebras we introduce
generalize corresponding commutative root space counterparts.
 In the special limit of tensor product power one and
commutative limit of a product in $E$, they
reduce to original continual Lie algebras.
In some cases mappings of noncommutative root space
continual Lie algebras we derive do not
form continual Lie algebras in commutative limit.

Many interesting properties  of continual Lie algebras have been
found useful in applications within the group-theoretical
approach~\cite{lezsav}  to the construction of exactly solvable dynamical systems
\cite{sv2, sv3, bakas1, bakas2, bakas3, zu0, zuz}.
 The structure of commutations relations and bilinear mappings~$K$
turned out to be very helpful in the development
 of noncommutative generalizations for known integrable models
\cite{zu0, zuz, zuz2}.
Continual Lie algebras with noncommutative root space we introduce
inspire various  new applications in integrable models. In particular,
they can be used (as in~\cite{zu0}) for construction of solvable
 models def\/ined in noncommutative spaces~\cite{sw, dim3, zuz}.

The plan of the paper is as follows. In Section~\ref{cla}
we give a formal def\/inition of a continual Lie algebra.
Then in Section~\ref{exc} examples of continual Lie algebras
with commutative root spaces  are mentioned.
In Section~\ref{exnc}
 we give def\/initions and notations related to noncommutative spaces of roots~(NCRS).
In Section~\ref{newx} we introduce new examples of mappings
in a noncommutative root space ${\cal E}$, and prove that they
satisfy to the def\/ining identities of a continual Lie algebra.
Finally, we mention directions for further generalizations and possible
applications in solvable models.

\section{Continual Lie algebras}
\label{cla}

Let ${\cal E}$ be a vector space.
 A continual Lie algebra \cite{s} is generated by
  the generalized local part
 ${\cal G}^{m_0}=\oplus_{|n| \le m_0} {\cal G}_n$,
 ${\cal G}_n=\{ X_n(\phi), \phi \in {\cal E}\}$ , $n \in \mathbb{Z}$,
    satisfying the def\/ining relations
 for all $\phi, \psi \in {\cal E}$, and $|n|, |m|, |n+m| \le m_0$,
\begin{equation}
\label{dr}
\left[X_n(\phi), X_m(\psi) \right]=X_{n+m}(K_{n,m}(\phi, \psi)),
\end{equation}
where $K_{n,m}: {\cal E} \times {\cal E} \to {\cal E}$,
$n$,$m \in \mathbb{Z}$, are bilinear mappings.
As for classical discrete root space Lie algebras,
we call ${\cal E}$ {\it the root space}.
   Ordinary Jacobi identity applied to  elements $X_i(\phi)$ imply the following
conditions on $K_{n,m}$:
\begin{gather}
\label{gk}
K_{k,m+n}(\phi, K_{ m,n}(\psi, \chi)) +
K_{m,n+k}(\psi, K_{ n,k}(\chi, \phi))
+ K_{ n,k+m}(\chi, K_{ k,m}(\phi, \psi))=0,
\\
\label{sk}
K_{ n,m}(\phi, \psi) = - K_{ m,n}(\psi, \phi),
\end{gather}
for all $\phi, \psi,  \chi \in {\cal E}$, and $|l|\le m_0$, where $l$ denotes
an index (or a sum of indexes) in~(\ref{gk}).
Then an inf\/inite-dimensional algebra ${\cal G}({\cal E}; K) = {\cal G}'({\cal E}; K)/J$
 is called a continual contragredient Lie algebra  where
${\cal G}'({\cal E}; K)$ is a Lie algebra freely generated by the minimal
(in accordance with~$m_0$) generalized local part ${\cal G}^{m_0}$, and $J$
is the largest homogeneous ideal with trivial intersection with~${\cal G}_0$
(consideration of the quotient is equivalent to imposing the Serre
relations in ordinary Lie algebra case)~\cite{sv2, sv3}.
  When $|m_0|\le 1$, the commutation relations (\ref{dr}) have the form
\begin{gather}
\notag
\left[ X_0(\phi), X_0(\psi) \right] = X_0(K_{0, 0}(\phi, \psi)),
\qquad
\left[ X_0(\phi), X_{\pm }(\psi) \right]  =
X_{\pm }(K_{ \pm }(\phi, \psi)),
\\
\left[ X_{+}(\phi), X_{-}(\psi) \right]  =  X_0(K_{ 0}(\phi, \psi)),\label{m0relations}
\end{gather}
 for all $\phi, \psi \in {\cal E}$,
and the conditions (\ref{gk}), (\ref{sk}) reduce to
\begin{gather}
\label{redjac1}
K_{0,0}(\phi, \psi)= - K_{0,0}(\psi, \phi),
\\
\label{redjac2}
K_{ \pm }(K_{ 0, 0}(\phi, \psi), \chi) =
K_{  \pm }(\phi, K_{ \pm }(\psi, \chi)) - K_{ \pm }
(\psi, K_{ \pm }
 (\phi, \chi)),
\\
\label{redjac3}
K_{ 0,0}( \psi, K_{ 0}(\phi, \chi)) =
K_{ 0}(K_{   +}(\psi, \phi), \chi) + K_{ 0}
(\phi, K_{  -}(\psi, \chi)),
\\
\label{redjac4}
K_{ 0,0}( \phi, K_{ 0,0}(\psi, \chi) ) +  K_{ 0,0} ( \psi,
K_{ 0,0}(\chi, \phi) ) +
K_{ 0,0}( \chi, K_{ 0,0}(\phi, \psi) ) =0.
\end{gather}

\section{Examples of continual Lie algebras}
\label{exc}

In this section ${\cal E}$ is a space of complex dif\/ferentiable functions.
Here we give examples of continual Lie algebras with commutative ${\cal E}$, relevant
to the constructions of this paper.
Other examples can be found in \cite{s, s10, sv2, sv3}.

\subsection{Witt algebra}

The Witt algebra \cite{ds} is the centerless Virasoro algebra.
The commutation relation on the single generator $X(\phi)$, for
$\phi$, $\psi \in {\cal E}$, are
\begin{gather}
\label{witt}
\left[X(\phi), X(\psi) \right]=X(\phi \partial \psi - \psi \partial \phi)
 \equiv X\big(\big[ \phi,^\partial \psi \big]\big),
\\
\label{witt1}
K(\phi, \psi)=\phi \partial \psi - \psi \partial \phi,
\end{gather}
where $\partial$ denotes the dif\/ferentiation with respect to a
real parameter with obvious notation after the last equality in (\ref{witt}).
  The only condition that the mapping
$K(\phi, \psi)$ satisf\/ies is (\ref{gk}) with $k=m=n=0$.

\subsection{Ricci f\/low algebra}

The Ricci f\/low algebra \cite{bakas1, bakas2, bakas3}
is determined by the bilinear mappings
\begin{gather}
K_{0,  0}(\phi, \psi) =  0,
\qquad
K_\pm (\phi, \psi)  =   \mp \phi \cdot \psi,
\qquad
K_0(\phi, \psi)  =   \partial (\phi \cdot \psi).\label{rfca1}
\end{gather}
The set of the mappings (\ref{rfca1}) is equivalent to the set
$  K_0(\phi, \psi)=  \phi \cdot \psi$,
$K_\pm (\phi, \psi) = \mp \phi \cdot \partial \psi$,
 $K_{0, 0}(\phi, \psi)= 0$.
It is easy to see that both sets obey the conditions (\ref{redjac1})--(\ref{redjac4}).  The
bicomplex construction \cite{bakas3, zuz2} based on generators of
this Lie algebra leads to the simplest example of the  Ricci f\/low
equation.

 \subsection{Poisson bracket algebra}

The third example we consider in this section is a continual Lie algebra
def\/ined by the mappings
\begin{gather}
K_{0,0}(\phi, \psi) = 0,
\qquad
K_{\pm }(\phi, \psi)  =  \mp i\partial \phi \cdot \psi,
\qquad
K_{0}(\phi, \psi) =  -i \partial(\phi \cdot \psi),\label{poissmap}
\end{gather}
and  $K_{n,m}(\phi, \psi)= i(n  \partial\psi \cdot \phi
 - m   \partial \phi \cdot  \psi)$, $n, m \in \mathbb{Z}$.
In \cite{sv2} it was proved that this continual Lie algebra
is isomorphic to the Poisson
bracket algebra under an appropriate substitution of variables.

\section{Noncommutative tensor product root space}
\label{exnc}

In all examples of continual Lie algebras  mentioned above
 ${\cal E}$ is necessarily a commutative space. This requirement is
dictated by the identities (\ref{redjac1})--(\ref{redjac4}).
 Indeed, to prove (\ref{redjac1})--(\ref{redjac4}) by
substituting corresponding mappings one should be able to
interchange elements of~${\cal E}$.
Certain problems could also be caused by
the presence of derivatives in mappings.
Note again that the general def\/inition of a continual Lie algebra admits
 a noncommutative space of roots, although almost all examples of
continual Lie algebras do not survive in noncommutativity.
  In order to overcome these dif\/f\/iculties
we def\/ine continual Lie algebras with
 mappings acting in the space of tensor product
 powers of a noncommutative space~$E$~\cite{at}.
We introduce counterparts of continual Lie algebras
discussed in previous section by def\/ining new mappings while keeping
the form of commutation relations~(\ref{m0relations}).
 Since the relations (\ref{redjac1})--(\ref{redjac4})
 for a continual Lie algebra mappings come from ordinary Jacobi identity
they are preserved in the root space  we use.

First we give some def\/initions and notations.
Let $E$ be a noncommutative algebra with a~product~$\cdot$.
 Then let ${\cal E}$ be the space of all tensor powers of $E$
(including possibly inf\/inite or semi-inf\/inite powers).
For simplicity we take identical copies of $E$ in ${\cal E}$,
 though a generalization with nonidentical $E$ is also  possible.
 We denote by
$E^{\otimes^M}=E\otimes \cdots \otimes E$,
 subspaces of ${\cal E}$ containing~$M$ copies of $E$.
For a monome $\phi \in {\cal E}$, which belongs to a subspace
$E^{\otimes^N}$ we def\/ine (possibly inf\/inite) ${\rm ord}\, \phi=N$.
 The product of two elements $\phi$, $\psi$,
 ${\rm ord}\; \phi= {\rm ord}\, \psi$, in the tensor algebra~${\cal E}$
 is def\/ined standardly (we skip the sign of the product)
$\phi \psi= \bigotimes\limits_{i=1}^{{\rm ord}\, \phi}
\phi_i \cdot \psi_i$, where
$\bigotimes\limits_{n} \eta_n$, denotes
the {\it ordered} tensor product.

Let us introduce two new operations.
 The ``gluing'' (or concatenation) operation $\widehat {}$
for two f\/inite  order elements $\phi, \psi \in {\cal E}$:
\begin{gather}
\notag
\phi \;  {\widehat {}}\;  \psi  =
\phi_1\otimes\cdots \otimes
 (\phi_{{\rm ord}\, \phi} \cdot \psi_1) \otimes \cdots \otimes
\psi_{{\rm ord}\, \psi}
\\
\phantom{\phi \;  {\widehat {}}\;  \psi}{}
=   \left( \bigotimes\limits_{n=1}^{{\rm ord}\, \phi-1} \phi_n \right)
\otimes (\phi_{{\rm ord}\, \phi} \cdot \psi_1) \otimes
  \left( \bigotimes\limits_{m=2}^{{\rm ord}\, \psi} \psi_m\right).\label{glu}
\end{gather}
The gluing operation is a mapping
$E^{\otimes^{{\rm ord}\, \phi}}
\times E^{\otimes^{{\rm ord}\, \psi}}
\longrightarrow E^{\otimes^{{\rm ord}\, \phi + {\rm ord}\, \psi-1}}$.
 The def\/inition of the gluing operation can be generalized
for the case of semi-inf\/inite tensor product elements. One can concatenate
left semi-inf\/inite with right semi-inf\/inite elements.

 Suppose $E$ possesses also a formal derivative operation $\partial$ (e.g., with respect to a
real para\-me\-ter) parameter.
In some cases we also assume the existence of the inverse (with respect to a product def\/ined in~$E$)
operator $\partial^{-1}$ to the formal derivative~$\partial$.
Then for $c\in E$, we def\/ine a~dif\/ferential operator
 $D_k$, $1 \le k \le {\rm ord} \, \phi$, which acts
on the ${\rm ord} \, \phi$ tensor power element
$\phi \in {\cal E}$, as follows
\begin{equation*}
D_k \phi = \phi_1 \otimes \cdots \otimes c\cdot \partial
 \phi_k \otimes \cdots \otimes \phi_{{\rm ord} \, \phi}
= \bigotimes\limits_{i=1}^{{\rm ord} \, \phi}
 (c\cdot \partial)^{\delta_{i,k}} \cdot \phi_i.
\end{equation*}
The gluing operation  (\ref{glu}), as well as the action of
$D_{ {\rm ord}\, \phi}
\left( \phi \;  {\widehat {}}  \; \psi\right)$, are
  associative with respect to the tensor product
\begin{gather*}
\phi \;  {\widehat {}}\;  (\psi \otimes \chi) =
(\psi \;  {\widehat {}}\; \psi) \otimes \chi,
\\
D_{ {\rm ord}\, \phi}(\phi \;  {\widehat {}}\;  (\psi \otimes \chi))
 =
( D_{ {\rm ord}\, \phi} (\phi \;  {\widehat {}}\; \psi)) \otimes \chi.
\end{gather*}
In what follows an element $c\in E$ is skipped for the sake of simplicity.

We introduce also the following operator in ${\cal E}$
\begin{equation}
\label{multpart}
{\partial }^{\otimes\,  \rm ord }\cdot \phi=
({\partial } \otimes \dots \otimes
 {\partial } )\cdot \phi
={\partial }^{\otimes^{\rm ord\, \phi}}\cdot \phi,
\end{equation}
with the number of derivatives ${\partial }$ in the tensor product
equal to the tensor power order of an element ${\cal E}$  which
${\partial }^{\otimes\,  \rm ord }\cdot$ acts on, e.g.,
${\rm ord} \, \phi$.
We  will use two notations for the action of the
dif\/ferentiation on an element $\phi \in {\cal E}$,
\begin{equation*}
  {\partial}^{\otimes\,  \rm ord }\cdot
 \phi \equiv \phi  \cdot  {\partial}^{\otimes\,  \rm ord }.
\end{equation*}
The natural property of the operator ${\partial }^{\otimes \, \rm ord}\cdot$ is
obvious
\begin{equation}
\label{natprop}
{\partial }^{\otimes \, \rm ord} \cdot (a\otimes b)
=
({\partial }^{\otimes \, \rm ord} \cdot a) \otimes
({\partial }^{\otimes \, \rm ord} \cdot b),
\end{equation}
where here and what follows the $\cdot$-multiplication has
higher priority with respect to the tensor product
  so that we will skip corresponding brackets.

 In (\ref{witt}) and what follows we denote by
$\left[\phi ,^{A}\psi\right]$ the commutator
\[
\left[\phi ,^{A} \psi\right]= \phi A \psi - \psi A \phi,
\]
where $A$ is an operation inserted in between $\phi$ and $\psi$.
We will also use the operator $P_\otimes$ which inverts the order of
tensor multipliers in an element of ${\cal E}$.
 Note that for $\phi, \psi \in {\cal E}$,
\begin{equation}
\label{pprop}
P_\otimes ( \phi\otimes \psi)=
P_\otimes ( \psi) \otimes P_\otimes ( \phi).
\end{equation}
 Having def\/ined mappings for new continual Lie algebras with noncommutative
spaces of roots~${\cal E}$ we have to establish
connections with their commutative counterparts, i.e., corresponding
continual Lie algebras with commutative root spaces.
In order to do so we consider the following commutative limit.
Firstly, we reduce~${\cal E}$ to~$E$ by taking the tensor order
of ${\cal E}$ equal to one. All tensor power product operations present
in def\/ining continual Lie algebra mappings $K$ have to be also replaced
with corresponding tensor product power one operations acting in~$E$.
Then we pass to a commutative ordinary product limit in all actions
of operators in~$E$ involving a~noncommutative $E$ product.
Alternatively one can consider another limit when all tensor products
have to be replaced by a noncommutative product in $E$,
and then pass to its commutative limit.
  In some cases in our constructions we do not assume that formal derivative operations
obey Leibniz rule with respect to a noncommutative product in~$E$
or tensor product in~${\cal E}$.
In the rest of the paper we proceed with examples of new continual Lie algebras
with noncommutative root spaces (NCRS)~${\cal E}$.

\section{New examples of NCRS continual algebras}
\label{newx}

\subsection{NCRS Witt algebra}

We start with the most trivial case.
One can guess the following mapping
\begin{gather}
\notag
K(\phi, \psi) =  \phi\; {  \widehat {}   }
\; D_1 \psi
- \psi\;  {  \widehat {}   } \; D_1 \phi
\\
\notag
 \phantom{K(\phi, \psi)}{} =
\phi_1\otimes \cdots \otimes (\phi_{{\rm ord}\, \phi}
\cdot \partial \psi_1)\otimes \cdots \otimes  \psi_{{\rm ord}\, \phi}
-
\psi_1\otimes \cdots \otimes (\psi_{{\rm ord}\, \psi}
 \cdot \partial \phi_1)\otimes \cdots \otimes  \phi_{{\rm ord}\, \phi}
\\
\label{ncwitt}
\phantom{K(\phi, \psi)}{} =  \big[ \phi,^{  \; \; {\widehat {} }
 \; \cdot  D_1   }
  \psi  \big].
\end{gather}
 We call a continual Lie algebra def\/ined by (\ref{ncwitt}) the {\it noncommutative root
space Witt} continual Lie algebra.

\begin{remark}
When ${\rm ord}\, \phi={\rm ord}\, \psi=1$ in~(\ref{ncwitt}),
the mapping formally coinsides with~(\ref{witt1}), though
the proof is still valid since~(\ref{gk}) has a dif\/ferent sense
(see~(\ref{wittproof})).
\end{remark}

\subsection{NCRS Ricci f\/low algebra}

Following the idea given in previous subsection we
introduce the following mappings $K_{0}$, $K_{\pm}$, $K_{0,0}$:
 \begin{gather}
\label{mapi1}
K_+( \phi ,  \psi  )  =
 \phi \otimes  \psi,
\\
\label{mapi2}
K_-( \phi , \psi  )  =
 - \psi \otimes  \phi,
\\
\label{mapi3}
K_{0, 0} (\phi , \psi  )  =
\big[\phi ,^\otimes
   \psi\big],
\\
\label{mapi4}
K_{0}(\phi, \psi)  =  D_{ {\rm ord}\; \phi}
\left( \phi \;  {\widehat {}}  \; \psi\right),
\end{gather}
where
\[
D_{ {\rm ord}\, \phi}
\left( \phi \;  {\widehat {}}  \; \psi\right) =
\phi_1\otimes\cdots \otimes
\partial(\phi_{{\rm ord} \, \phi} \cdot \psi_1) \otimes \cdots
\otimes \psi_{{\rm ord}\, \psi}.
\]
We call a continual Lie algebra def\/ined by the mappings
 (\ref{mapi1})--(\ref{mapi4}) the {\it noncommutative root space Ricci flow}
continual Lie algebra.
The relation (\ref{redjac1}) is hold in
$({\rm ord}\, \phi + {\rm ord}\, \psi)$-tensor power of~$E$,~(\ref{redjac2}),~(\ref{redjac4}) are hold in
$({\rm ord}\, \phi + {\rm ord}\, \psi+
{\rm ord}\, \chi)$-power, while
the relation (\ref{redjac3}) is in
$(  {\rm ord}\, \phi + {\rm ord}\, \psi+
{\rm ord}\, \chi -1)$.
In the commutative limit, we put ${\rm ord}\, {\cal E} =1$, substitute
all tensor product remaining in mappings by the product in $E$,
and then consider $E$ being commutative.  Then we see that
the mappings (\ref{mapi1})--(\ref{mapi4}) have the commutative
limit~(\ref{rfca1}).

\subsection{NCRS Poisson bracket algebra}

The case of the Poisson bracket algebras seems to be
more complicated.
We introduce a continual Lie algebra over a noncommutative
root space so that its mappings satisfy to the identities
\mbox{(\ref{redjac1})--(\ref{redjac4})}, and comply with a commutative limit~(\ref{poissmap}). Consider
\begin{gather}
\label{ncpbp}
K_{+} (\phi, \psi) =  - i\,
{\partial }^{\otimes \, {\rm ord} }  \cdot \phi \otimes \psi,
\\
\label{ncpbm}
K_{-} (\phi, \psi)  =  i \, \psi \otimes
{\partial }^{\otimes \, {\rm ord } }
  \cdot \phi,
\\
\label{ncpb00}
K_{0, 0} (\phi, \psi)  =
 - i \, \big[\phi {,}^\otimes \psi\big],
\\
\label{ncpb0}
K_{0} (\phi, \psi)
 =
-i \, (\partial^{-1})^{\otimes \, {\rm ord } }\cdot
   D^2_{{\rm ord } \, \phi}\cdot (\phi\; \widehat{}\;  \psi),
\end{gather}
where the operator $(\partial^{-1})^{\otimes \, {\rm ord } }\,\cdot$ acts similar to
(\ref{multpart}).

\begin{remark}
Note that $K_{\pm}$, $K_{0,0}$  (\ref{ncpbp})--(\ref{ncpb0})
 have the Poisson bracket continual Lie algebra mappings as a commutative limit.
As specif\/ied above, we put ${\rm ord}\, {\cal E} =1$, substitute
all tensor product remaining in mappings by the product in $E$,
and then consider a commutative limit of the  product in $E$.
Then
 $ K_{ 0, 0}(\phi, \psi)=0$,
and
$K_{ 0}(\phi, \psi)= -i \partial\,(\phi \cdot \psi)$,
$K_{\pm}(\phi, \psi)= \mp i \partial\phi \cdot \psi$.
\end{remark}

\begin{remark}
The mappings def\/ining the higher grading subspaces
of the Ricci f\/low continual Lie algebra~(\ref{rfca1}), as well as for the Ricci f\/low
(\ref{mapi1})--(\ref{mapi4}) and Poisson bracket (\ref{ncpbp})--(\ref{ncpb0})
 continual Lie algebras with noncommutative root spaces
will be discussed in a separate paper.
\end{remark}

\subsection{Further examples}
\label{fe}

In this subsection we give further examples of noncommutative
root space continual Lie algeb\-ras whose mappings do not form
continual Lie algeb\-ras with commutative root spaces in
the commutative limit.

\subsubsection{NCRS Poisson-type bracket continual Lie algebras}

For $\phi,\psi \in {\cal E}$, consider the mappings:
\begin{gather}
\label{ncppb0}
  K_0(\phi, \psi) =
- i\;  D_{{\rm ord} \, \phi} (\phi \;{\widehat {}} \; \psi),
\\
\label{ncppbp}
  K_{+} (\phi, \psi) =  -i  \,
  \phi \;{\widehat {}} \; (D_{1} \psi),
\\
\label{ncppbm}
  K_{-} (\phi, \psi) =  i  \,  \psi \;{\widehat {}} \; (D_{1} \phi),
\\
\label{ncppb00}
  K_{0,0} (\phi, \psi) =   K_+(\phi, \psi) + K_-(\phi, \psi)
= - i \, \big[\phi ,^{\; {  \widehat {}   } \cdot D_{1} } \psi\big].
\end{gather}
 In the commutative limit, ${\rm ord}\, {\cal E}=1$,
and commutative product in $E$, the mappings (\ref{ncppb0})--(\ref{ncppb00}) reduce to
$K_{0}(\phi, \psi) =  -i \partial( \phi \psi)$,
$K_{+}(\phi, \psi) =  - i \phi \partial \psi$,
$K_{-}(\phi, \psi) =  i \psi \partial \phi$,
$K_{0,  0}(\phi, \psi) = -i \left[ \phi,^{\partial} \psi \right]$.
 It is easy to check that this set of mappings does
not satisfy to (\ref{redjac1})--(\ref{redjac4}),
 and therefore does not def\/ine a continual Lie algebra.
We see that the form of the mapping $K_{-}$ (\ref{ncppbm})
represents a~direct noncommutative analog of
the Poisson bracket continual Lie algebra mapping $K_{-}$ of~(\ref{poissmap})
with the commutative root space, while $K_{+}$-mapping (\ref{ncppbp})
contains the derivative action on the second argument in contrast
to $K_{+}$ of~(\ref{poissmap}).

\begin{remark}
One can alternatively def\/ine a continual Lie algebra described above
 by the mappings:
\begin{gather*}
  K_0(\phi, \psi) =
- i/2\,( D_{{\rm ord}\, \phi} (\phi \;{\widehat {}} \; \psi)
+  D_{{\rm ord} \, \psi} (\psi \;{\widehat {}} \; \phi)) ,
\\
  K_{+} (\phi, \psi) =   -i  \,
(D_{{\rm ord} \, \phi} \phi) \;{\widehat {}} \; \psi,
\\
  K_{-} (\phi, \psi) =  i  \,
(D_{{\rm ord} \, \psi} \psi) \;{\widehat {}} \; \phi,
\\
  K_{0,0} (\phi, \psi) =  K_+(\phi, \psi) + K_-(\phi, \psi),
\label{ncmpb00}
\end{gather*}
that in the commutative limit provide
$K_{0}(\phi, \psi)=- i\partial(\phi \psi)$,
$K_{+} (\phi, \psi)= -i (\partial \phi)\psi $,
$K_{-} (\phi, \psi)=i (\partial \psi) \phi $,
$K_{0,0} (\phi, \psi)=i ( (\partial \psi)\phi - (\partial \phi)\psi)$,
and, as in (\ref{ncppb0})--(\ref{ncppb00}), do not correspond
 a continual Lie algebra with a commutative root space.
We call these two examples the {\it Poisson-type bracket} continual Lie algebras.
\end{remark}

\subsubsection{NCRS Integral mapping continual Lie algebra}

Let $E$ be the ring of dif\/ferentiable functions with a noncommutative
product $\cdot$ and dif\/feren\-tia\-tion~$\partial$. Then let ${\cal E}$ be the
algebra of tensor powers of~$E$.
We then introduce the mappings of a~new continual Lie algebra with
noncommutative root space~${\cal E}$ and the mapping~$K_0$ of integral type
\begin{gather}
\label{ncintp}
K_{+} (\phi, \psi)  =   - i \big[
 \phi \cdot {\partial }^{\otimes \, {\rm ord} }
  \otimes  \psi
+\psi \otimes
   {\partial }^{\otimes \, {\rm ord } }
   \cdot P_{\otimes} \phi \big],
\\
\label{ncintm}
K_{-} (\phi, \psi)  =   i \big[  (P_{\otimes}\,
 \phi) \cdot    {\partial }
^{\otimes \, {\rm ord } }
 \otimes \psi
+\psi \otimes  {\partial }^{\otimes \, {\rm ord } }
  \cdot \phi
\big],
\\
\label{ncint00}
K_{0, 0} (\phi, \psi) =
 - i   \big[\phi {,}^\otimes \psi\big],
\\
\label{ncint0}
K_{0} (\phi, \psi) =  - i \,(\partial^{-1})^{\otimes \, {\rm ord}} \cdot (\phi \otimes \psi).
\end{gather}
The terms containing $P_\otimes$-operator
 in (\ref{ncintp}) and (\ref{ncintm})
are not important for the construction of corresponding continual Lie algebra
(see the proof below) and are not present in commutative case.
  It is a usual situation since noncommutative counterparts for derivative terms
are not unique in general.
In contrast to the case of the noncommutative
Poisson bracket continual Lie algebra the derivative acts on the second
argument in $K_+(\phi, \psi)$.

\begin{remark}
In the commutative limit the mappings $K_{\pm}$, $K_{0,0}$
 (\ref{ncintp})--(\ref{ncint0}) reduce to $K_{ 0, 0}(\phi, \psi)=0$,
$K_{ 0}(\phi, \psi)= -i \partial^{-1} (\phi \cdot \psi)$,
$K_{\pm}(\phi, \psi)= \mp i ( \phi \cdot\partial \psi  + \psi \cdot\partial \phi)$,
and do not form a continual Lie algebra.
\end{remark}

\begin{remark}
In a similar fashion, we can introduce noncommutative counterparts for
other examples of continual Lie algebras (in particular for vector f\/ields and dif\/feomorphisms, and
cross-product continual Lie algebras \cite{sv2})
 with commutative ${\cal E}$ described in \cite{s, sv3}.
The case of contact Lie algebras \cite{sc, sc1}
 with noncommutative root spaces will be considered elsewhere.
\end{remark}

\subsection{Main statement}

Now we will show that new examples of mappings acting on noncommutative root spaces
in subsections above do indeed comply with the def\/inition of a continual Lie algebra.

\begin{proposition}
\label{prowitt}
The mappings
 $a)$~\eqref{ncwitt},
$b)$~\eqref{mapi1}--\eqref{mapi4},
$c)$~\eqref{ncpbp}--\eqref{ncpb0},
$d)$~\eqref{ncppb0}--\eqref{ncppb00},
$e)$~\eqref{ncintp}--\eqref{ncint0}
satisfy to the identities \eqref{redjac1}--\eqref{redjac4} and define
 noncommutative root space continual Lie algebras.
\end{proposition}

The cases $a)$, $b)$, $c)$ represent
generalizations for the Witt, Ricci f\/low, and
Poisson bracket continual Lie algebras correspondingly.

\begin{proof}
$a)$ The relation (\ref{redjac1}) is satisf\/ied trivially.
 For (\ref{redjac4}) one has
\begin{gather}
\nonumber
 K( \phi, K(\psi, \chi) ) +  K ( \psi, K(\chi, \phi) ) +
K( \chi, K(\phi, \psi) )
\\
\label{wittproof}
\qquad {} =\phi\; \widehat{} \;D_1 \psi\; \widehat{}\; D_1 \chi
- \psi\; \widehat{} \; D_1  \chi\;  \widehat{} \;  D_1 \phi
-
\phi\; \widehat{}\; D_1 \chi\;  \widehat{} \; D_1 \psi
+
\chi \; \widehat{} \; D_1 \psi\;  \widehat{} \; D_1 \phi
 \\
\nonumber
 \qquad  \quad {} +
\mbox{\rm two   permutations   of} \  (\phi, \psi, \chi)
 =0.
\end{gather}
Then we see that all terms in the above expression
cancel.

$b)$
(\ref{redjac1}) and (\ref{redjac4}) trivially follow from the
def\/inition (\ref{mapi3}) of $K_{0,0}$.
 The identity (\ref{redjac2}) follows
from the def\/initions of $K_{0,0}$ (\ref{mapi3}) and $K_\pm$ (\ref{mapi1})--(\ref{mapi2}).
For (\ref{redjac3}) one has for any
$\phi, \psi, \chi \in {\cal E}$,
\begin{gather*}
 K_{ 0,0}( \psi, K_{ 0}(\phi, \chi))
 =
\left[\psi ,^\otimes  \phi_1 \otimes \cdots \otimes
\partial ( \phi_{{\rm ord}\, \phi} \cdot \chi_1)\otimes \cdots \otimes
\chi_{{\rm ord}\, \chi}\right]
\\
 \qquad = D_{{\rm ord}\; \phi}(
\psi \otimes    \phi_1\otimes \cdots \otimes
 \phi_{{\rm ord}\, \phi} \cdot \chi_1 \otimes \cdots
\otimes \chi_{{\rm ord}\, \chi} )
\\
 \qquad \quad {}-  D_{{\rm ord}\, \phi}( \phi_1\otimes \cdots \otimes
 \phi_{{\rm ord}\, \phi } \cdot \chi_1
\otimes \cdots \otimes \chi_{{\rm ord}\, \chi} \otimes \psi)
\\
  \qquad {} =
K_0(\psi \otimes \phi, \chi) - K_0(\phi, \chi \otimes \psi)
\\
  \qquad =
K_{ 0}(K_{   +}(\psi, \phi), \chi) + K_{ 0}
(\phi, K_{  -}(\psi, \chi)).
\end{gather*}

$c)$ The identities (\ref{redjac1}) and
(\ref{redjac4}) are trivially satisf\/ied by $K_{0,0}$  (\ref{ncpb00}).
Then we check (\ref{redjac2}) for (\ref{ncpbp}) and (\ref{ncpb00}):
\begin{gather*}
 K_{  +}(\phi, K_{ +}(\psi, \chi)) - K_{ + }
(\psi, K_{ +} (\phi, \chi))
\\
  \qquad {} =
  - ({\partial }^{\otimes \, \rm ord} \cdot \phi) \otimes
({\partial }^{\otimes \, \rm ord} \cdot \psi) \otimes
 \chi
 +
({\partial }^{\otimes \, \rm ord}\cdot  \psi) \otimes
({\partial }^{\otimes \, \rm ord} \cdot \phi) \otimes
 \chi
\\
 \qquad {} = -i \,
{\partial }^{\otimes \, \rm ord} \cdot
\left(-i \big[\phi,^{\otimes} \psi\big] \right)\otimes \chi
\\
 \qquad {}= K_{+}(K_{ 0, 0}(\phi, \psi), \chi),
\end{gather*}
and similarly for $K_-$ (\ref{ncpbm}). Here
we have made use of the property~(\ref{natprop}).
Next we check (\ref{redjac3})
\begin{gather*}
K_{ 0}(K_{   +}(\psi, \phi), \chi) + K_{ 0}
(\phi, K_{  -}(\psi, \chi))
 =
K_{ 0}(-i\partial^{\otimes \, \rm ord} \cdot \psi \otimes \phi, \chi) +
K_{ 0}
(\phi, i \chi \otimes \partial^{\otimes \, \rm ord} \cdot \psi)
\\
\qquad{}=
  - (\partial^{-1})^{\otimes \, {\rm ord } }\cdot
\big[
\partial^{\otimes \, \rm ord} \cdot \psi \otimes
D^2_{{\rm ord }\, \phi}\cdot (\phi\; \widehat{}\;  \chi)
-
D^2_{{\rm ord } \, \phi}\cdot (\phi\; \widehat{}\;  \chi)
\otimes \partial^{\otimes \, \rm ord} \cdot \psi
\big]
\\
 \qquad {} =
  - \big[\psi ,{}^{\otimes}\,
(\partial^{-1})^{\otimes \, {\rm ord } }\cdot
   D^2_{{\rm ord }\, \phi}\cdot (\phi\; \widehat{}\;  \chi)\big]
\\
\qquad {} =K_{ 0,0}( \psi, -i  (\partial^{-1})^{\otimes \, {\rm ord } }\cdot
   D^2_{{\rm ord } \, \phi}\cdot (\phi\; \widehat{}\;  \chi))
\\
 \qquad {} =
K_{ 0,0}( \psi, K_{ 0}(\phi, \chi)).
\end{gather*}

$d)$
As we showed in the proof of $a)$,
(\ref{ncppb00}) satisfy (\ref{redjac1}) and (\ref{redjac4}).
Then proving (\ref{redjac2}) one has
\begin{gather*}
  K_+(K_{0,0}(\phi, \psi), \chi)
  =  -  \phi_1 \otimes \cdots \otimes
(\phi_{{\rm ord} \, \phi} \cdot \partial \psi_1)
 \otimes \cdots \otimes (\psi_{{\rm ord} \, \psi}
\cdot \chi_1)\otimes \cdots \chi_{{\rm ord} \, \chi}
\\
\phantom{K_+(K_{0,0}(\phi, \psi), \chi)=}{}
 + \psi_1\otimes \cdots \otimes (\psi_{{\rm ord} \, \psi}
 \cdot \partial \phi_1)\otimes \cdots
\otimes(\phi_{{\rm ord} \, \phi} \cdot \chi_1) \otimes \cdots
\chi_{{\rm ord} \, \chi}
\\
\phantom{K_+(K_{0,0}(\phi, \psi), \chi)}{}
= K_+ (\phi, -i \psi_1\otimes \cdots
\otimes(\psi_{{\rm ord} \, \psi}
 \cdot \partial \chi_1)\otimes \cdots \otimes\chi_{{\rm ord} \, \chi})
\\
\phantom{K_+(K_{0,0}(\phi, \psi), \chi)=}{}
 - K_-(\psi,  -i \phi_1 \otimes\cdots \otimes
(\phi_{{\rm ord} \, \phi} \cdot \partial \chi_1)
\otimes \cdots \otimes\chi_{{\rm ord} \, \chi}),
  \end{gather*}
and similar for $K_-(\phi, \psi)$  in  (\ref{redjac2}).
Proving (\ref{redjac3}) we f\/ind
\begin{gather*}
  K_{0,0}(\psi,  - i \phi_1\otimes\cdots  \otimes
\partial(\phi_{{\rm ord} \, \phi} \cdot \chi_1)
\otimes \cdots \otimes\chi_{{\rm ord} \, \chi})
\\
  \qquad {}=  - \psi_1 \otimes  \cdots  \otimes
 (\psi_{{\rm ord} \, \psi} \cdot \partial \phi_1)
\otimes  \cdots  \otimes \partial
(\phi_{{\rm ord} \, \phi} \cdot \chi_1) \otimes \cdots
\otimes \chi_{{\rm ord} \, \chi}
\\
   \qquad \quad {}+
 \phi_1 \otimes \cdots  \otimes \partial
(\phi_{{\rm ord} \, \phi} \cdot \partial \chi_1)\otimes \cdots
 \otimes (\chi_{{\rm ord} \; \chi} \cdot \psi_1)\otimes \cdots
\otimes \psi_{{\rm ord} \, \psi}
\\
  \qquad {}=
K_0(-i \psi_1\otimes  \cdots
\otimes (\psi_{{\rm ord} \, \psi} \cdot \phi_1)\otimes \cdots , \chi)
\\
  \qquad \quad {}+
K_0(\phi, i \chi_1 \otimes \cdots  \otimes
(\chi_{{\rm ord} \, \chi} \cdot \psi_1)\otimes \cdots
\otimes\psi_{{\rm ord} \, \psi}).
\end{gather*}

$e)$ The identities (\ref{redjac1}) and
(\ref{redjac4}) are trivially satisf\/ied by $K_{0,0}$  (\ref{ncint00}).
We then check (\ref{redjac2}) for (\ref{ncintp}) and (\ref{ncint0})
\begin{gather*}
 K_{  +}(\phi, K_{ +}(\psi, \chi)) - K_{ + }
(\psi, K_{ +} (\phi, \chi))
\\
  \qquad{} =
  - (\phi \cdot {\partial }^{\otimes \, \rm ord} )  \otimes
(   \psi\cdot {\partial }^{\otimes \, \rm ord}
\otimes  \chi +
   \chi \otimes  {\partial}^{\otimes \, \rm ord} \cdot
 P_\otimes \psi)
\\
\qquad \quad {}-
(   \psi\cdot    {\partial }^{\otimes \, \rm ord}
 \otimes  \chi +
   \chi \otimes   {\partial}^{\otimes \, \rm ord} \cdot
   P_\otimes \psi)
\otimes ((P_\otimes\, \phi) \cdot  {\partial }^{\otimes \, \rm ord})
\\
\qquad \quad {}+
(\psi \cdot   {\partial }^{\otimes \, \rm ord} )  \otimes
(   \phi\cdot {\partial }^{\otimes \, \rm ord}
\otimes  \chi +
   \chi \otimes   {\partial}^{\otimes \, \rm ord} \cdot
 P_\otimes \phi)
\\
 \qquad \quad {}+
(   \phi\cdot {\partial }^{\otimes \, \rm ord}
 \otimes  \chi +
   \chi \otimes   {\partial}^{\otimes \, \rm ord} \cdot
   P_\otimes \phi)
\otimes ((P_\otimes\, \psi) \cdot  {\partial }^{\otimes \, \rm ord})
\\
 \qquad {}=
 - ( \phi \cdot {\partial }^{\otimes \, \rm ord} \otimes
  (\psi \cdot {\partial }^{\otimes \, \rm ord})
-
  \psi \cdot {\partial }^{\otimes \, \rm ord} \otimes
  (\phi \cdot {\partial }^{\otimes \, \rm ord})) \otimes \chi
\\
\qquad \quad{} -
 \chi \otimes (( {\partial}^{\otimes \, \rm ord}
\cdot  P_\otimes \psi)
\otimes
    ((P_\otimes \, \phi)
 \cdot {\partial }^{\otimes \, \rm ord} )
+
(  {\partial }^{\otimes \, \rm ord}
  \cdot P_\otimes \phi) \otimes
    ((P_\otimes \, \psi) \cdot {\partial }^{\otimes \, \rm ord} )
)
\\
\qquad {}= -i(
 - i \big[ \phi {,}^\otimes \psi\big] \cdot {\partial }^{\otimes \, \rm ord}
 \otimes \chi
- i
\chi \otimes P_\otimes  {\partial}^{\otimes \, \rm ord} \cdot
 \left[\phi {,}^\otimes \psi\right])
\\
 \qquad {}= K_{+}(K_{ 0, 0}(\phi, \psi), \chi),
\end{gather*}
and similarly for the mapping $K_-$~(\ref{ncintm}). Here
we have used the properties~(\ref{natprop}) and~(\ref{pprop}).
 We then check the identity (\ref{redjac3}) for (\ref{ncintp})--(\ref{ncint0}):
\begin{gather*}
K_{ 0}(K_{   +}(\psi, \phi), \chi) + K_{ 0}
(\phi, K_{  -}(\psi, \chi))
\\
 \qquad {}=
 - (\partial^{-1})^{\otimes \, {\rm ord}}\cdot\big (
\big[
 \psi \cdot {\partial }^{\otimes \, \rm ord}
 \otimes \phi
+ \phi \otimes  {\partial }
^{\otimes \, \rm ord}
 \cdot P_{\otimes} \psi)
 \big] \otimes \chi\big)
\\
\qquad\quad {} +
 (\partial^{-1})^{\otimes \, {\rm ord}} \cdot \big( \phi \otimes
\big[(P_{\otimes} \;\psi) \cdot  {\partial }^{\otimes \, \rm ord}
   \otimes \chi
+\chi \otimes  {\partial }^{\otimes \, \rm ord} \cdot \psi
 \big]\big)
\\
\qquad{}=
  - (\partial^{-1})^{\otimes \, {\rm ord}}\cdot (\psi \cdot {\partial }^{\otimes \, \rm ord}
\otimes (\phi \otimes \chi))
 + (\partial^{-1})^{\otimes \, {\rm ord}} \cdot  ((\phi \otimes \chi)
\otimes {\partial }^{\otimes \, \rm ord} \cdot \psi )
\\
 \qquad {} =
 - \psi  \otimes  (\partial^{-1})^{\otimes \, {\rm ord}} \cdot (\phi \otimes \chi)
 +   (\partial^{-1})^{\otimes \, {\rm ord}} \cdot (\phi \otimes \chi)\otimes \psi
\\
 \qquad {} =
K_{ 0,0}( \psi, - i (\partial^{-1})^{\otimes \, {\rm ord}} \cdot (\phi\otimes  \chi))
\\
 \qquad {} =
K_{ 0,0}( \psi, K_{ 0}(\phi, \chi)).\tag*{\qed}
\end{gather*}
\renewcommand{\qed}{}
\end{proof}

\section{Outlook}

In this paper we construct new examples of continual Lie
algebras with noncommutative root spaces.
Our considerations are focused on the Witt, Ricci f\/low, and
Poisson bracket continual Lie algebras that contain derivative
operations in the def\/ining mappings.
 When a root space ${\cal E}$
is noncommutative, derivative terms cause problems in f\/inding
solutions to the identities  (\ref{redjac1})--(\ref{redjac4}) following from
 Jacobi identity.
To f\/ix this problem and introduce generalization for the above mentioned
continual Lie algebras we have chosen ${\cal E}$
to be the space of tensor product powers of a space $E$ with a
noncommutative product.

The actions of mappings we use in the space of tensor product powers
can be seen as linear part approximations for more general
construction of bilinear mappings in ${\cal E}$.
Even at the this level, the examples of continual
 Lie algebras introduced in this paper can be generalized.
In particular one can use
more general mappings containing arbitrary operators $D_i$,
$i \in 1, \dots, {\rm ord }\, \phi$, $\phi \in {\cal E}$, as well
as more complicated dif\/ferential mappings.
 Relations to \cite{berenret} as well as interesting generalizations
will be discussed elsewhere.
We propose to def\/ine continual super Lie algebras,
 $q$-deformations (as in \cite{ds})  of the noncommutative continual
Witt algebra, and further develop  $q$-deformed counterparts
\cite{reshsem} for (noncommutative) continual Lie algebras.
It would be also interesting to determine possible Hopf algebra
structure associated to continual Lie algebras described here.
In a noncommutative root space continual Lie algebra construction
 we use two types of products: a noncommutative product $\cdot$ in a
 space $E$, and the tensor product def\/ining the
 algebra ${\cal E}$ of tensor powers of $E$.
Thus it would be interesting to make connections with \cite{dim3} where
integrable systems have been constructed in spaces with two noncommutative multiplications.

Continual Lie algebras with noncommutative root spaces
 appear to be attractive objects both from algebraic point of view and
 in applications in integrable models.  In particular, these algebras can
def\/ine generalizations of certain exactly solvable models.
Integrable models def\/ined in noncommutative spaces
 with ordinary Lie algebras as an algebraic origin were constructed in
 \cite{dim3, gri, ham1, zuz2}.
Using the bicomplex construction \cite{zuz} for continual Lie
algebras with noncommutative root spaces one derives associated dynamical
systems in noncommutative spaces.
Finally, one can try to def\/ine vertex operators \cite{zusav} as well as vertex
operator algebras associated to continual Lie algebras
with noncommutative root spaces.

\subsection*{Acknowledgements}
This work was supported by Science Foundation of Ireland Frontiers
of Research Grant.
We would like also to thank V.~Kac, A.~Perelomov, A.~Rosenberg, and M.~Tuite
 for valuable discussions.

\pdfbookmark[1]{References}{ref}
\LastPageEnding
\end{document}